\documentclass[twocolumn]{aastex62}

\usepackage{graphicx}
\usepackage{stmaryrd}
\usepackage{multirow}
\usepackage{natbib}
\usepackage{xspace}
\usepackage{hyperref}
\usepackage{amsmath}
\usepackage{longtable}

\newcommand{\simi}{\ensuremath{\sim}}

\newcommand{\degree}{\ensuremath{^{\circ}}\xspace}
\newcommand{\xos}{\mbox{XO-7}\xspace}
\newcommand{\xosb}{\mbox{XO-7 b}\xspace}
\newcommand{\tess}{\textit{TESS}\xspace}
\newcommand{\kepler}{\textit{Kepler}\xspace}
\newcommand{\corot}{\textit{CoRoT}\xspace}
\newcommand{\jwst}{\textit{JWST}\xspace}
\newcommand{\hst}{\textit{HST}\xspace}

\newcommand{\modif}{}

\providecommand{\bjdtdb}{\ensuremath{\rm {BJD_{TDB}}}}

\providecommand{\mj}{\ensuremath{{\rm M_{\rm J}}}\xspace}
\providecommand{\rj}{\ensuremath{\rm R_{\rm J}}\xspace}

\providecommand{\fave}{\langle F \rangle}
\providecommand{\fluxcgs}{10$^9$ erg s$^{-1}$ cm$^{-2}$}

\newcommand{\vsini}{\ensuremath{v\,{\rm sin}\,i}\xspace}

\shorttitle{XO-7 \MakeLowercase{$\rm b$}}

\shortauthors{Crouzet et al.}

\begin{document}

\title{XO-7 b: A transiting hot Jupiter with a massive companion on a wide orbit}

\correspondingauthor{Nicolas Crouzet}
\email{nicolas.crouzet@esa.int}

\author[0000-0001-7866-8738]{Nicolas Crouzet}
\affiliation{Science Support Office, Directorate of Science, European Space Research and Technology Centre (ESA/ESTEC), Keplerlaan 1, 2201 AZ Noordwijk, The Netherlands}

\author[0000-0002-7718-7884]{Brian F. Healy}
\affiliation{Department of Physics and Astronomy, Johns Hopkins University, 3400 North Charles Street, Baltimore, MD 21218, USA}

\author{Guillaume H\'ebrard}
\affiliation{Institut d'Astrophysique de Paris, UMR7095 CNRS, Universit\'e Pierre \& Marie Curie, 98bis boulevard Arago, 75014 Paris, France}

\author{P. R. McCullough}
\affiliation{Department of Physics and Astronomy, Johns Hopkins University, 3400 North Charles Street, Baltimore, MD 21218, USA}
\affiliation{Space Telescope Science Institute, 3700 San Martin Dr, Baltimore, MD 21218, USA}

\author{Doug~Long}
\affiliation{Space Telescope Science Institute, 3700 San Martin Dr, Baltimore, MD 21218, USA}

\author{Pilar~Monta\~n\'es-Rodr\'iguez}
\affiliation{Instituto de Astrof\'isica de Canarias, C. V\'ia L\'actea s/n, E-38205 La Laguna, Tenerife, Spain}
\affiliation{Universidad de La Laguna, Dept. de Astrof\'isica, E-38206 La Laguna, Tenerife, Spain}

\author{Ignasi Ribas}
\affiliation{Institut de Ci\`encies de l'Espai (CSIC-IEEC), Campus UAB, Carrer de Can Magrans s/n, 08193 Bellaterra, Spain}
\affiliation{Institut d'Estudis Espacials de Catalunya (IEEC), Gran Capit\`a, 2-4, Edif. Nexus, 08034 Barcelona, Spain}

\author{Francesc Vilardell}
\affiliation{Institut de Ci\`encies de l'Espai (CSIC-IEEC), Campus UAB, Carrer de Can Magrans s/n, 08193 Bellaterra, Spain}
\affiliation{Institut d'Estudis Espacials de Catalunya (IEEC), Gran Capit\`a, 2-4, Edif. Nexus, 08034 Barcelona, Spain}

\author{Enrique~Herrero}
\affiliation{Institut de Ci\`encies de l'Espai (CSIC-IEEC), Campus UAB, Carrer de Can Magrans s/n, 08193 Bellaterra, Spain}
\affiliation{Institut d'Estudis Espacials de Catalunya (IEEC), Gran Capit\`a, 2-4, Edif. Nexus, 08034 Barcelona, Spain}

\author{Enrique~Garcia-Melendo}
\affiliation{Serra H\'unter Fellow, Escola Superior d'Enginyeries Industrial, Aeroespacial i Audiovisual, Universitat Polit\`ecnica de Catalunya, Carrer Colom 1, 08222 Terrasa, Spain}

\author{Matthieu Conjat}
\affiliation{Laboratoire Lagrange, Universit\'{e} C\^ote d'Azur, Observatoire de la C\^ote d'Azur, CNRS, Boulevard de l'Observatoire, CS 34229, F-06304 Nice Cedex 4, France}

\author{Jerry Foote}
\affiliation{Vermillion Cliffs Observatory, 4175 E. Red Cliffs Drive, Kanab, UT 84741, USA}

\author{Joe Garlitz}
\affiliation{Elgin Observatory, Elgin, OR, USA}

\author{Phillip Vo}
\affiliation{Department of Physics and Astronomy, University of Waterloo, Waterloo, ON, N2L 3G1, Canada}

\author{Nuno C. Santos}
\affiliation{Instituto de Astrof\'isica e Ci\^encias do Espa\c{c}o, Universidade do Porto, CAUP, Rua das Estrelas, 4150-762 Porto, Portugal}
\affiliation{Departamento de F\'isica e Astronomia, Faculdade de Ci\^encias, Universidade do Porto, Rua do Campo Alegre, 4169-007 Porto, Portugal}
          
\author{Jos de Bruijne}
\affiliation{Science Support Office, Directorate of Science, European Space Research and Technology Centre (ESA/ESTEC), Keplerlaan 1, 2201 AZ Noordwijk, The Netherlands}

\author{Hugh P. Osborn} 
\affiliation{Aix Marseille Universit\'e, CNRS, CNES, LAM (Laboratoire d'Astrophysique de Marseille), 13388 Marseille, France}

\author{Shweta Dalal} 
\affiliation{Institut d'Astrophysique de Paris, UMR7095 CNRS, Universit\'e Pierre \& Marie Curie, 98bis boulevard Arago, 75014 Paris, France} 

\author{Louise D. Nielsen}
\affiliation{Observatoire de Gen\`eve,  Universit\'e de Gen\`eve, 51 Chemin des Maillettes, 1290 Sauverny, Switzerland}

\begin{abstract}

Transiting planets orbiting bright stars are the most favorable targets for follow-up and characterization. We report the discovery of the transiting hot Jupiter XO-7 b and of a second, massive companion on a wide orbit around a circumpolar, bright, and metal rich G0 dwarf (V = 10.52, $T_{\rm eff} = 6250 \pm 100 \; \rm K$, $\rm[Fe/H] = 0.432 \pm 0.057 \; \rm dex$). We conducted photometric and radial velocity follow-up with a team of amateur and professional astronomers. XO-7 b has a period of $ 2.8641424 \pm 0.0000043$ days, a mass of $0.709 \pm 0.034 \; \mj$, a radius of $1.373 \pm 0.026 \; \rj$, a density of $0.340 \pm 0.027 \; \rm g \, cm^{-3}$, and an equilibrium temperature of $1743 \pm 23 \; \rm K$. Its large atmospheric scale height and the brightness of the host star make it well suited to atmospheric characterization. The wide orbit companion is detected as a linear trend in radial velocities with an amplitude of $\sim100 \; \rm m \, s^{-1}$ over two years, yielding a minimum mass of $4 \; \mj$; it could be a planet, a brown dwarf, or a low mass star. The hot Jupiter orbital parameters and the presence of the wide orbit companion point towards a high eccentricity migration for the hot Jupiter. Overall, this system will be valuable to understand the atmospheric properties and migration mechanisms of hot Jupiters and will help constrain the formation and evolution models of gas giant exoplanets.

\end{abstract}

\keywords{Stars: planetary systems --- Planets and satellites: individual (\xosb) --- Methods: observational --- Techniques: photometric --- Techniques: spectroscopic}


\section{Introduction} 
\label{sec: Introduction}

Gas giant planets transiting bright stars on a close-in orbit are favorable targets for detailed studies. They can be detected and followed-up in photometry and radial velocity, and their atmosphere can be observed by spectroscopy. Ground-based surveys with small apertures and wide fields of view \modif{such as WASP \citep{Pollacco2006, Collier2007}, HATNet \citep{Bakos2004}, HATSouth \citep{Bakos2013}, KELT \citep{Pepper2007}, QES \citep{Alsubai2013} discovered most of the hot Jupiters known to date including \simi90 around relatively bright stars ($V < 11$)}. The \corot \citep{Baglin2009} and \kepler \citep{Borucki2010} missions detected a few such systems but they targeted mostly fainter stars. The \tess mission \citep{Ricker2015} is an all sky survey and should detect nearly all hot Jupiters transiting stars of magnitude $I < 13$ \citep{Sullivan2015, Sullivan2017}.

The presence of wide orbit companions in hot Jupiter systems draws particular interest. 
Two mechanisms have been proposed to bring gas giant planets to close-in orbits: disk migration or high eccentricity migration. The latter requires a high initial eccentricity, potentially due to scattering by another massive companion. These  migration mechanisms should in principle be reflected in the orbital parameters of hot Jupiters \citep{Faber2005}. 
Besides, wide orbit companions may affect the orbit of planets that are closer to the star in the form of an exchange between eccentricity and inclination via the Lidov-Kozai mechanism \citep{Lidov1962, Kozai1962}. This mechanism has been investigated to explain the eccentricity and obliquity distributions of hot Jupiters. 
However, no correlation has been found between misaligned or eccentric hot Jupiters and the frequency of massive companions on wide orbits \citep{Knutson2014, Ngo2015, Piskorz2015, Ngo2016}. \modif{In their sample of 51 planets, \citet{Knutson2014} find statistically significant accelerations in 15 systems and derive an occurrence rate of $51\% \pm 10\%$ for companions with masses between 1--13 \mj and orbital semi-major axes between 1--20~au. To date, ten systems with a transiting hot Jupiter and a massive, well characterized, wide orbit planetary companion are known (HAT-P-13, HAT-P-17, HAT-P-44, HAT-P-46, HATS-59, HD 219134, KELT-6, WASP-41, WASP-47, WASP-134)\footnote{Source: \url{http://exoplanet.eu/}}. Discovering and characterizing such systems will help shed light on the formation and orbital evolution of gas giant exoplanets.}

The XO project \citep{McCullough2005} aims at detecting transiting exoplanets around bright stars from the ground with small telescopes. The project started in 2005 and discovered five close-in gas giant planets, XO-1b to XO-5b \citet{McCullough2006, Burke2007, JohnsKrull2008, McCullough2008, Burke2008}. A second version of XO was deployed in 2011 and 2012 and operated from 2012 to 2014. This led to the discovery of XO-6 b, a hot Jupiter transiting a fast rotating F5 star on an oblique orbit \citep{Crouzet2017}. In this paper, we report the discovery of \xosb, a transiting hot Jupiter orbiting a bright G0V star with a massive companion on a wide orbit.
We present the instrumental setup and data reduction used to detect the transiting object in Sec.~\ref{sec: XO photometry}, and describe the follow-up campaign by amateur and professional astronomers to characterize the system in Sec.~\ref{sec: Follow-up campaign}. The analysis of these data is detailed in Sec.~\ref{sec: Analysis}. The XO-7 system properties are given in Sec.~\ref{sec: System parameters} and discussed in Sec.~\ref{sec: Discussion}. Conclusions are given in Sec.~\ref{sec: Conclusion}.

\section{XO photometry}
\label{sec: XO photometry}

The second version of XO consists of three identical units located at Vermillion Cliffs Observatory, Kanab, Utah, at Observatorio del Teide, Tenerife, Canary Islands, and at Observatori Astron\`omic del Montsec (OAdM), Sant Esteve de la Sarga, Spain. Each unit is composed of two 10 cm diameter and 200 mm focal length Canon telephoto lenses equipped with an Apogee E6 $1024 \times 1024$ pixels CCD camera and an R band filter, mounted on a German-Equatorial Paramount ME mount and protected by a shelter with a computer-controlled roof. Each unit operates robotically. The six lenses and cameras operate in a network configuration and point towards the same fields of view, which do not overlap with those of the original XO survey. The CCDs are used in Time Delayed Integration (TDI): pixels are read continuously while stars move along columns on the detector. The recorded images are long strips of $43.2^{\circ} \times 7.2^{\circ}$. This technique maximises the number of observed bright stars and increases the observing efficiency. The exposure time is 5.3 minutes for a full strip and the nominal Point Spread Function (PSF) Full Width Half Maximum (FWHM) is 1.2 pixels. We observed two strips starting from the north celestial pole and descending along RA 6h and 18h over two separate nine-month periods \modif{between 2012 and 2014.}

We carved the strips into $1024 \times 1024$ pixel images, which yields 9 fields of $7.2^{\circ} \times 7.2^{\circ}$ \modif{with a pixel scale of 25.3 arcsec per pixel}. We performed the astrometry using the astrometry.net software program\footnote{\url{http://astrometry.net/}} \citep{Lang2010} followed by a 6 parameter astrometric solution. Science frames are calibrated with darks and flat fields, which are 1-dimensional arrays for TDI images, and corrected for warm columns. We ran circular aperture photometry using the Stellar Photometry Software program \citep{Janes1993} with an aperture size optimized as a function of stellar magnitude. We implemented several photometric calibrations and built lightcurves for the 2000 brightest stars in each square field of view (up to $V = 12$). We removed systematic effects using the Sysrem algorithm \citep{Tamuz2005}, combined the lightcurves from the six cameras, searched for periodic signals using the Box Least Square (BLS) algorithm \citep{Kovacs2002}, and kept the signals that were compatible with planetary transits for visual inspection. More details on the instrumental setup, instrumental performances, and data reduction procedure can be found in \citet{McCullough2005, Crouzet2017, Crouzet2018}.

\begin{figure}[htbp]
   \centering
   \includegraphics[width=\columnwidth]{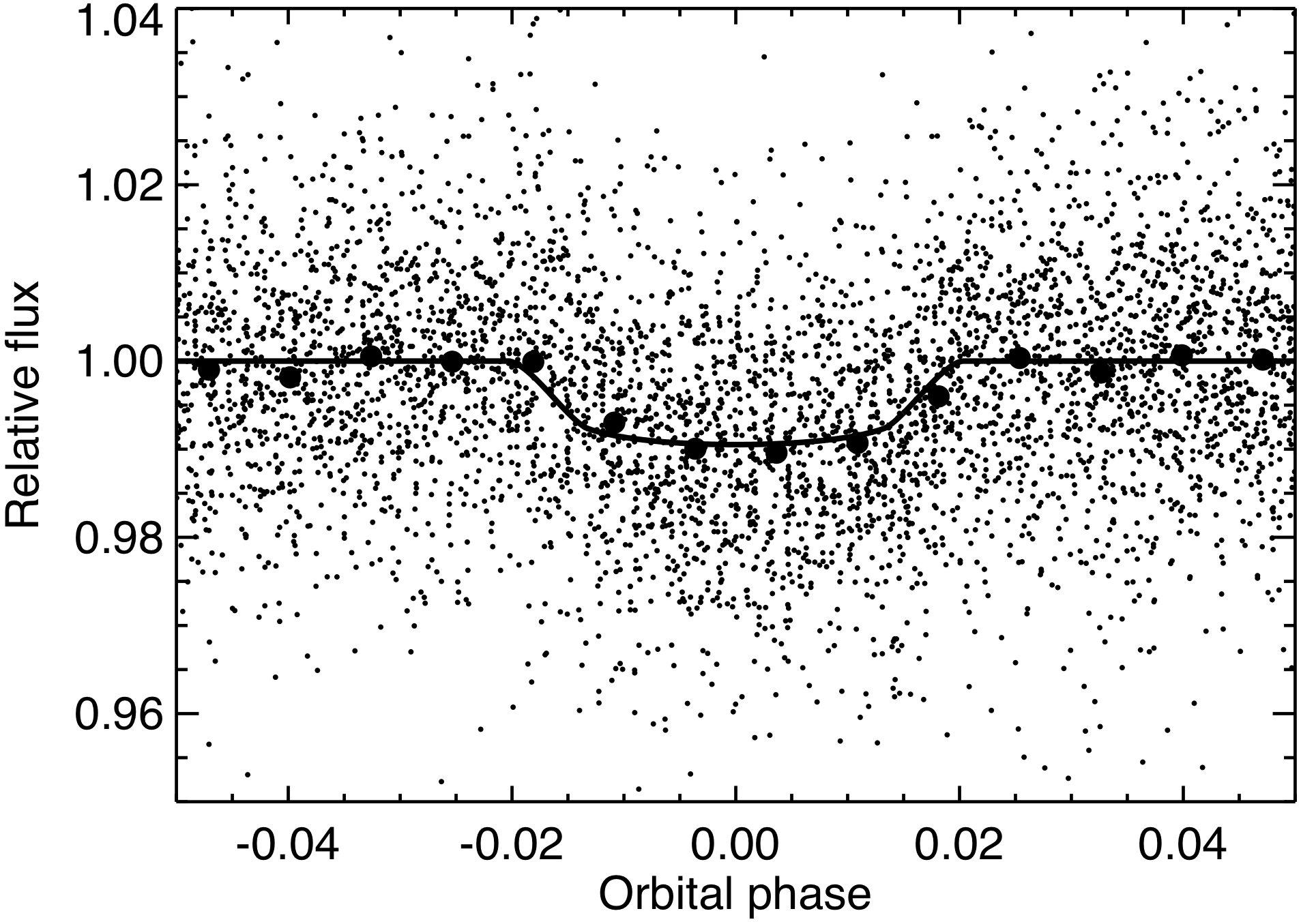}
      \caption{Phase-folded discovery lightcurve of \xosb showing \modif{the original data (black dots), the data binned over 30-minute intervals (black filled circles)}, and the best transit fit using the parameters from Table~\ref{tab: system parameters} (black line).}
   \label{fig: xo lightcurve}
\end{figure}

The phase-folded discovery lightcurve of \xosb is shown in Fig.~\ref{fig: xo lightcurve}. The host star is relatively bright (BD+85 317, V = 10.52, see Table \ref{tab: stellar parameters}). \modif{We gathered 43880 exposures of this object between September 28, 2012 and June 7, 2014.} The lightcurve dispersion calculated using an outlier resistant estimate over the out-of-transit data points is 1.1\%. After binning the lightcurve over a timescale of 30 min, the dispersion is 0.6\%. After phasing the lightcurve at the planet's orbital period and binning over 30 min (137 phase bins), the dispersion drops to 0.09\% (900 ppm) owing to the large quantity of collected data. The transit seen in the discovery lightcurve motivated an extensive follow-up campaign to characterize this system.

\section{Follow-up campaign}
\label{sec: Follow-up campaign}

\subsection{Faint nearby star}
\label{sec: Faint nearby star}

\begin{figure}[htbp]
   \centering
   \includegraphics[width=3.5cm]{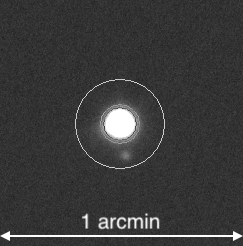}
   \caption{Example of an image cropped around the planet host star XO-7 (BD+85 317) taken during the photometric follow-up with the 40 cm Schaumasse telescope at the Observatoire de Nice, France. The background K dwarf is located 8 arcsec away from the planet host star and is well resolved. The circular aperture and the annulus used to measure the stellar flux and the sky background respectively are shown.}
   \label{fig: image Schaumasse}
\end{figure}

The planet host star has a nearby star at a separation of 8" that is five magnitudes fainter ($G$ = 15.8407, where $G$ is the Gaia G band magnitude). This neighbour is a K star ($T_{\rm eff} = 4038\; \rm K$, $BP-RP = 1.87$, where $BP$ and $RP$ are magnitudes from the Gaia blue and red photometers), it has a parallax of $2.9315 \pm 0.0354$ mas, a proper motion of $0.947 \pm 0.066$ and $17.027 \pm 0.082$ mas/yr in $RA$ and $Dec$ respectively, and an estimated distance of $338 \pm 4$ pc as inferred from Gaia DR2 \citep{Gaia2018, BailerJones2018}. It is located about 140 pixels in the scan direction on the Gaia detectors and does not affect the astrometry of the main target (which could be the case for separations of 10-20 pixels). Its Gaia DR2 astrometric data do not show anything suspicious. The parallax, proper motion, and distance of the main target are reported in Table~\ref{tab: stellar parameters}. These measurements show that both stars are unbound and the K star is in the background. They are not resolved by the XO instruments but the K star is faint enough to be negligible at the level of precision of the XO data. Both stars are well resolved in the follow-up observations (Fig. \ref{fig: image Schaumasse}). No other companion with a magnitude difference less than five is present within one arcmin in the bands used for the detection and follow-up (from B to i').

\subsection{Photometric follow-up}

\begin{figure}[htbp]
   \centering
   \includegraphics[width=8.5cm]{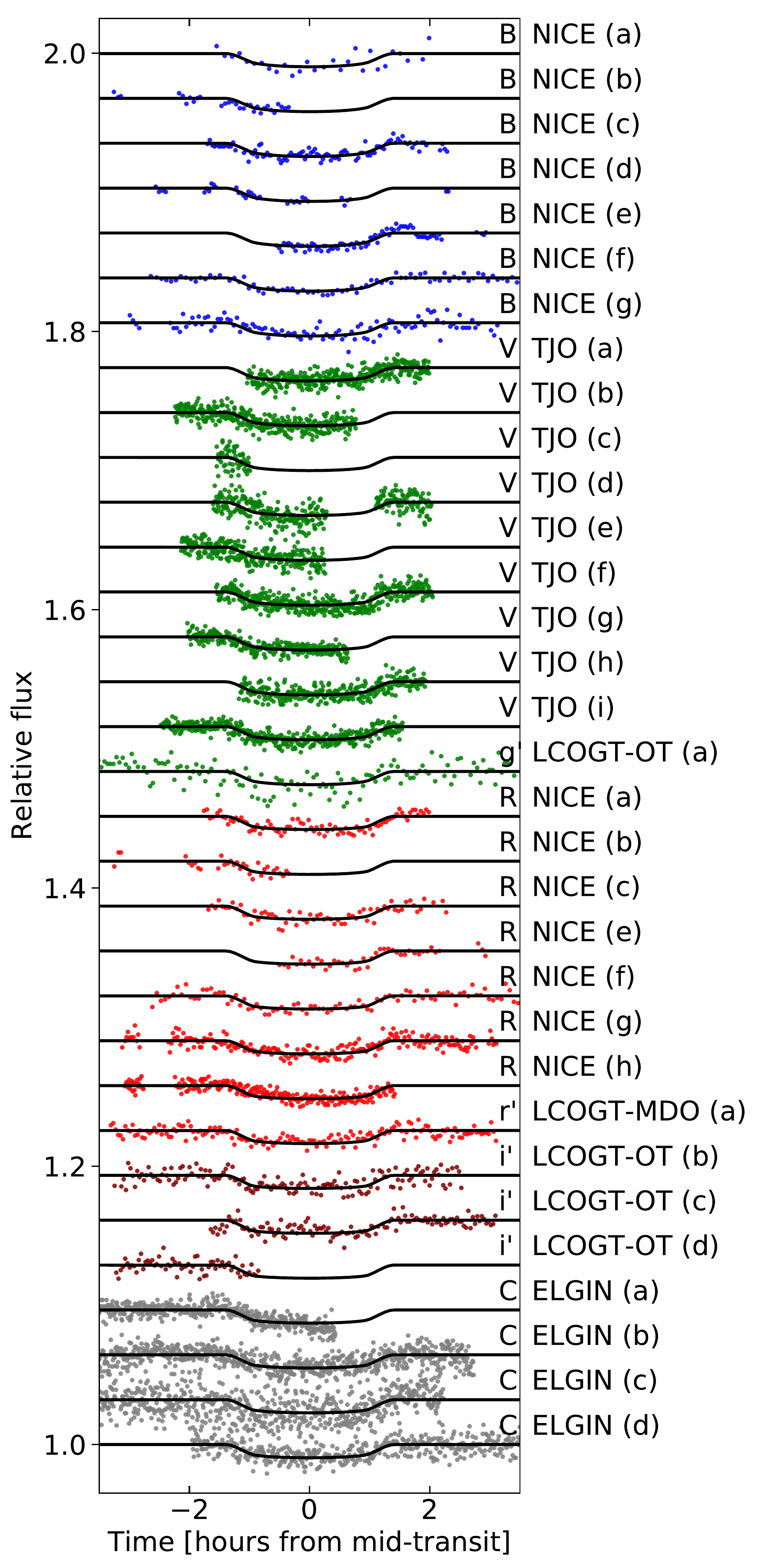}
   \caption{Photometric follow-up of XO-7 b. Individual transits are displayed. Bandpasses are noted B (blue), V and g' (green), R and r' (red), i' (dark red), and C (grey); observatories and observation dates are labeled as in Table \ref{tab: photometric follow-up}. The best transit model calculated in each bandpass is overplotted as a black line. The lightcurves do not have the same time sampling; thus, the apparent point-to-point dispersion is not representative of their relative quality. Lightcurves are offset for clarity.}
   \label{fig: lc individual transits}
\end{figure}

\begin{figure}[htbp]
   \centering
   \includegraphics[width=9cm]{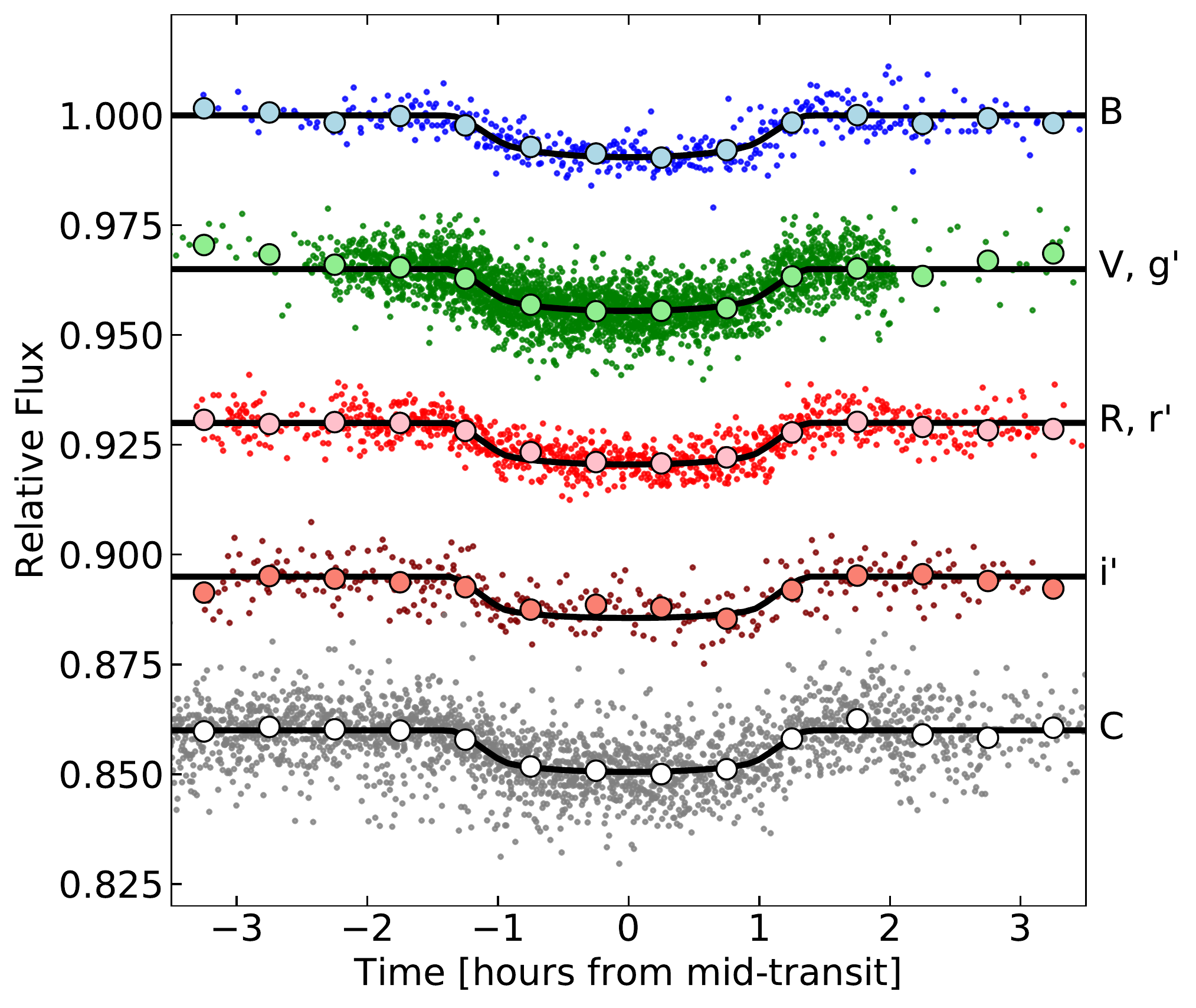}
      \caption{Photometric follow-up observations of XO-7 b gathered by bandpasses: B (blue), V and g' (green), R and r' (red), i' (dark red), and C (grey) from top to bottom. Data from different observations are blended in their respective filter bands. The best transit model in each bandpass is overplotted as a black line. We also plot the median flux values of bins spaced by 30 minutes for each lightcurve. Lightcurves are offset for clarity.}
   \label{fig: lc filters}
\end{figure}

\modif{Extensive photometric follow-up was conducted by a team of amateur and professional astronomers. We observed 22 transit events between June 12, 2017 and December 14, 2018 with facilities reported in Table \ref{tab: photometric follow-up} using different filters (B, V, $\rm g^\prime$, R, $\rm r^\prime$, $\rm i^\prime$) or without a filter (labelled as C for ``Clear''). We obtained 32 good quality lightcurves that we used in the analysis (Figs. \ref{fig: lc individual transits} and \ref{fig: lc filters}).}

\modif{Observations with the 40 cm Schaumasse telescope at the Observatoire de Nice (Nice, France) were conducted using a Johnson B or R filter, sometimes alternating between the two. The images were calibrated using bias, darks, and flat fields. We reduced the data with the IRIS astronomical image processing software \citep{Buil2005} and performed differential aperture photometry with four reference stars chosen from their brightness and photometric stability.}
\modif{Observations with the 80 cm Telescopi Joan Or\'o telescope at Montsec Astronomical Observatory (Lleida, Spain) were conducted with the MEIA2 instrument, a $2 \rm k \times 2 \rm k$ Andor CCD camera with a pixel scale of 0.36 arcsec and a squared field of view of 12.3 arcmin, using a Johnson V filter. The images were calibrated with darks, bias, and flat fields with the ICAT pipeline \citep{Colome2006}. Differential photometry was extracted with AstroImageJ \citep{Collins2017} using the aperture size and the set of comparison stars that minimised the rms of the out-of-transit photometry.} 
\modif{Observations with the 30 cm telescope at Elgin Observatory (Elgin, Oregon, USA) were conducted using a CCD without a filter. The images were calibrated using bias, darks, and flat fields. We reduced the data with the AIP4Win v2.4.8 Magnitude Measurement Tool \citep{Berry2005} and performed differential aperture photometry with three or five reference stars depending on the image and seeing quality. These stars were selected for each data set based on lowest noise and lack of curvature in the lightcurve and were averaged.}
\modif{Observations with the 40 cm telescopes of the Las Cumbres Observatory Global Telescope network (LCOGT) were conducted using Sloan Digital Sky Survey (SDSS) $\rm g^\prime$, $\rm r^\prime$, and $\rm i^\prime$ filters. The images were reduced with the LCOGT's BANZAI pipeline \citep{McCully2018} including bad-pixel masking, bias and dark subtraction, flat fielding, and image plate solving from astrometry.net \citep{Lang2010}. We performed aperture photometry using the Astropy Photutils package \citep{bradley2019}. Changes in the target star’s position on the detector between nights led us to use nine reference stars for the $\rm g^\prime$ band and ten for the $\rm r^\prime$ and $\rm i^\prime$ bands.}

\modif{We gathered these lightcurves using a consistent format and analysed them jointly. We converted the dates into Barycentric Julian Date (BJD) and performed a transit fit for each bandpass, during which each lightcurve was corrected for a linear trend and outliers more than three sigma away were removed. We used these consistent, corrected lightcurves for the combined fit in Section \ref{sec: Combined fit}.}

\begin{table*}
\begin{center}
\caption{Facilities used for the photometric follow-up.}
\label{tab: photometric follow-up}
\begin{tabular}{lll}
\hline
\hline
Observatory & Telescope & Label \\
\tableline 
Observatoire de Nice, France & Schaumasse, 16 inches (40 cm)  & NICE \\
\multicolumn{1}{@{~~~~~~~}l}{(a) 2017-06-12, (b) 2017-07-02, (c) 2017-07-25, (d) 2017-08-14} & FOV: 31'$\times$23'; Pixel size: 0.56"/px \\
\multicolumn{1}{@{~~~~~~~}l}{(e) 2017-08-17, (f) 2017-09-06, (g) 2018-08-07, (h) 2018-08-27} \\
Observatori Astron\`omic del Montsec, Catalonia, Spain & Joan Or\'o Telescope, 31 inches (80 cm)  &  TJO \\
\multicolumn{1}{@{~~~~~~~}l}{(a) 2017-06-12, (b) 2017-07-02, (c) 2017-09-06, (d) 2017-10-14} & FOV: 12.3'$\times$12.3'; Pixel size: 0.36"/px \\
\multicolumn{1}{@{~~~~~~~}l}{(e) 2017-10-16, (f) 2017-11-26, (g) 2018-08-05, (h) 2018-08-07} \\
\multicolumn{1}{@{~~~~~~~}l}{(i) 2018-12-14} \\
Elgin Observatory, Elgin, Oregon, USA & 12 inches (30 cm)  &  ELGIN \\
\multicolumn{1}{@{~~~~~~~}l}{(a) 2017-06-24, (b) 2017-07-17, (c) 2017-08-06, (d) 2017-08-09}  & FOV: 15.7'$\times$10.5'; Pixel size: 1.23"/px \\
Las Cumbres Observatory, McDonald Observatory, TX, USA  &  16 inches (40 cm)  &  LCOGT-MDO  \\
\multicolumn{1}{@{~~~~~~~}l}{(a) 2018-06-20} & FOV: 29'$\times$19'; Pixel size: 0.57"/px \\
Las Cumbres Observatory, Observatorio del Teide, Tenerife, Spain  &  16 inches (40 cm)   &  LCOGT-OT  \\
\multicolumn{1}{@{~~~~~~~}l}{(a) 2018-06-25, (b) 2018-08-07, (c) 2018-08-30, (d) 2018-09-19} & FOV: 29'$\times$19'; Pixel size: 0.57"/px \\
\hline
\hline
\end{tabular}
\end{center}
\vspace{-1mm}
Notes. Letters beneath each observatory indicate the dates of observation (cf. Fig. \ref{fig: lc individual transits}). The field of view  (FOV) in arcminute and pixel scale in arcsecond are also indicated.
\end{table*}

\subsection{Radial velocity follow-up}
\label{sec: Radial velocity follow-up}

\begin{figure}[htbp]
   \centering
    \includegraphics[width=8cm]{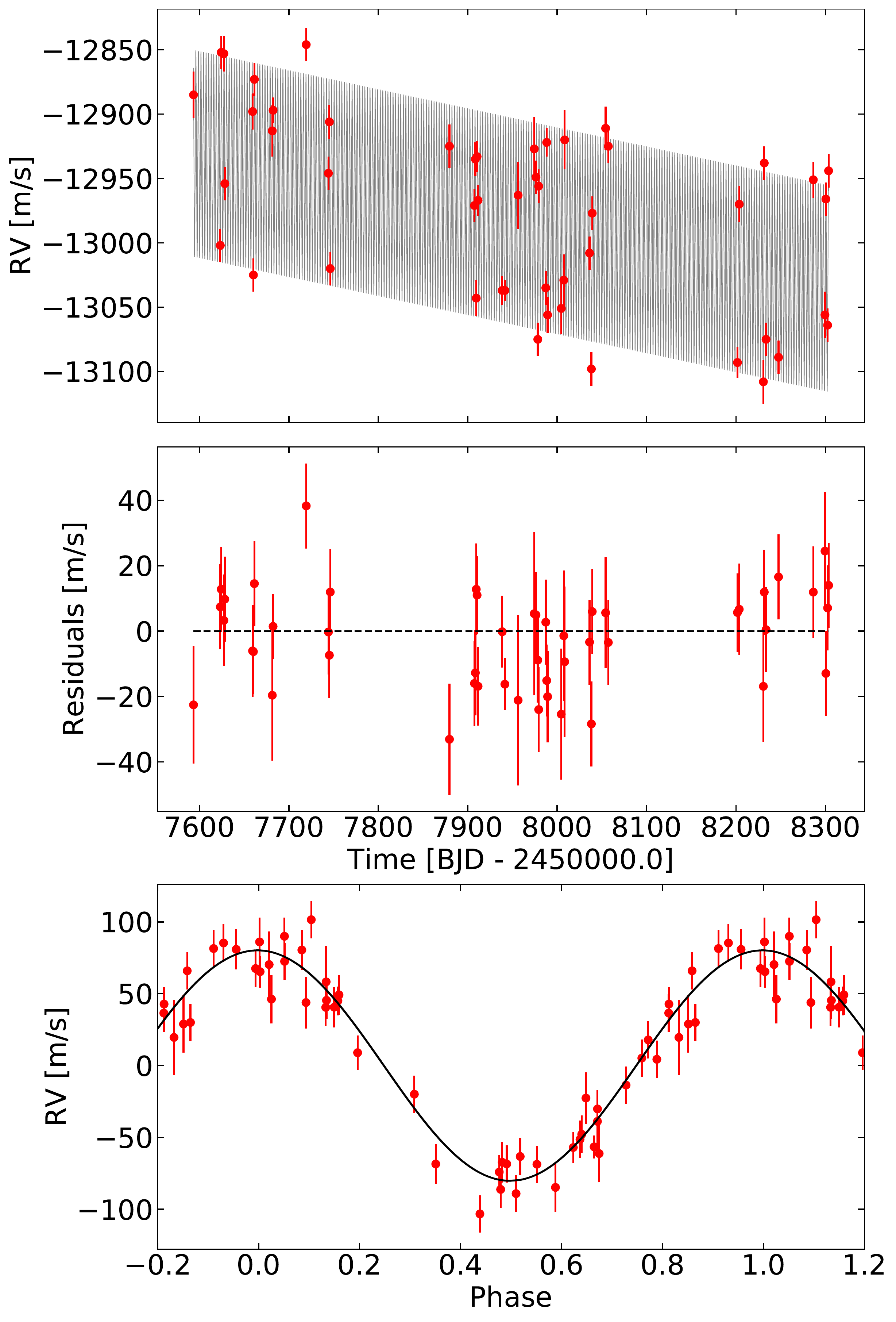}
      \caption{SOPHIE radial velocities of \xos with 1$\sigma$ error bars (red). In the top panel, the grey area represents a circular Keplerian fit to the hot Jupiter-induced motion combined with a linear trend for the unseen companion. The middle panel shows the residuals. In the bottom panel, we plot the RVs and model fit with the linear trend subtracted, phase-folded at the hot Jupiter's orbital period.}
   \label{fig: xo7 RVs}
\end{figure}

Radial velocity (RV) measurements were obtained between July 23, 2016 and July 4, 2018 with the SOPHIE spectrograph \citep{Perruchot2008, Bouchy2009, Bouchy2013} at the 193-cm telescope of Observatoire de Haute-Provence, France (Fig. \ref{fig: xo7 RVs}, Appendix \ref{tab: xo7 RVs}). We used its High-Resolution mode (resolving power $R=75\,000$). Exposure times were around 13 minutes allowing signal-to-noise ratios of around 27 per pixel at 550 nm to be reached on most of the exposures. We used the SOPHIE pipeline to extract the spectra from the detector images, cross-correlate them with a numerical mask which produces clear cross-correlation functions (CCFs), then fit the CCFs by Gaussians to derive the RVs \citep{Baranne1996, Pepe2002}. 

The resulting CCFs have a contrast that represents $\sim31$\,\%\ of the continuum, and a full width at half maximum (FWHM) of 11.0\,km/s showing some stellar rotation (we measured \vsini$=6 \pm 1 $~km/s from the CCF width; see Section \ref{sec: Spectral analysis of the host star}). The RVs have typical uncertainties around $\pm13$~m/s, whereas we removed from our final dataset three exposures having uncertainties larger than $\pm30$~m/s. Only five spectra were contaminated by moonlight. We estimated and corrected for that contamination by using the second SOPHIE fiber aperture, which was placed on the sky while the first aperture pointed toward the target \citep[e.g.][]{Hebrard2008, Bonomo2010}; this resulted in RV corrections around 35~m/s or smaller (whereas the dispersion of the residuals after the combined fit in Sec.\ref{sec: Combined fit} is 14.7~m/s). Excluding those five Moon-contaminated observations does not significantly change our results. The final RV dataset shows significant variations in phase with the transit ephemeris and with a semi-amplitude around 80~m/s implying a companion mass in the giant-planet regime, as shown in Fig. \ref{fig: xo7 RVs}. 

Radial velocities measured using different stellar masks (G2, K0, or K5) produce variations with similar amplitudes, so it is unlikely that these variations are produced by blend scenarios composed of stars of different spectral types. We finally adopted the RVs obtained with the K0 mask as they provide the least dispersed residuals. Using the RVs obtained from the G2 mask does not significantly change our results. Similarly, the measured CCF bisector spans quantify possible shape variations of the spectral lines. They show no correlations with the RVs, and no significant variations: their dispersion is two times smaller than the RV dispersion whereas each bisector span is roughly half as precise as the corresponding RV measurement. This reinforces the conclusion that the RV variations are due to a planetary companion, and not caused by spectral-line profile changes attributable to blends or stellar activity.

\section{Analysis}
\label{sec: Analysis}

\subsection{Spectral analysis of the host star}
\label{sec: Spectral analysis of the host star}

We begin our analysis of the data with a study of the host star. Stellar atmospheric parameters ($T_{\rm eff}$, $\log{g}$ and [Fe/H]) and respective uncertainties were derived using the methodology described in \citet{Sousa2008} and \citet{Santos2013}. In brief, we make use of the equivalent widths of tens of iron lines and we assume ionization and excitation equilibrium. The process makes use of a grid of Kurucz model atmospheres \citep{Kurucz1993} and the radiative transfer code MOOG \citep{Sneden1973}. 

The equivalent widths were measured on a SOPHIE spectrum built from the addition of the spectra used for the RV measurements, but excluding the five SOPHIE spectra presenting moonlight contamination. We obtained $T_{\rm eff} = 6220 \pm 70$\,K, log\,$g = 4.2 \pm 0.1$ (cgs), and [Fe/H] = $+0.48 \pm 0.05$. Using the calibration of \citet{Torres2010} with a correction following \citet{Santos2013}, we derive a mass and radius of $1.43 \pm 0.09$\,$\mathrm{M}_{\odot}$ and $1.47 \pm 0.20$\,$\mathrm{R}_{\odot}$, respectively. We also derived the projected rotational velocity \vsini$=6 \pm 1 $~km/s from the parameters of the CCF using the calibration of \citet{Boisse2010}.

\subsection{Combined fit}
\label{sec: Combined fit}

Proceeding to a comprehensive analysis of the system, we fit the photometric follow-up lightcurves and the radial velocities together using EXOFASTv2 \citep{Eastman2019}. In each lightcurve, we verified that the uncertainties of individual data points were of the same order of the standard deviation of the lightcurve (after subtracting the transit), to ensure that they were not under- or overestimated. In some cases, we rescaled the uncertainties accordingly. We used MIST stellar isochrones \citep{Dotter2016,Choi2016,Paxton2011,Paxton2013,Paxton2015}, an SED constructed from Tycho \citep{Hog2000}, 2MASS \citep{Cutri2003,Skrutskie2006} and WISE \citep{Cutri2014} catalog magnitudes, and the Gaia DR2 parallax \citep{Gaia2018} for BD+85 317 to constrain the host star's parameters. We use a quadratic limb darkening law:
\begin{equation}
\frac{I_{\mu}}{I_0} = 1 - u_1\,(1 - \mu) - u_2\,(1 - \mu)^2
\end{equation}
where $I$ is the intensity and $\mu$ is the angle between a line normal to the stellar surface and the line of sight of the observer. The limb-darkening coefficients $u_1$ and $u_2$ are free parameters with theoretical values interpolated from updated \citet{Claret2017} tables. In the fit, we set four priors: we used the SOPHIE spectral analysis to constrain the effective temperature and metallicity for the stellar isochrone fitting. We also set a prior on the Gaia parallax with an updated uncertainty (see Sec. \ref{sec: Analysis of Gaia DR2 data}.) Finally, we used \citet{Schlafly2011} measurements to place an upper limit on  V-band extinction \citep[see also][]{Green2019}. We set unconstrained starting values (e.g. transit depth and duration) based on the posteriors of a brief fitting run. In the full MCMC, we simultaneously fit a model to the transits and RVs, including a linear fit to the long-term trend from the latter data. The chains were well-mixed (Gelman-Rubin statistic \textless\ 1.01) after $\sim$ 37,000 steps.

The final parameter values and their uncertainties are computed as the medians and 1-sigma values of their respective posterior distribution functions. The limb darkening coefficients are free and are treated in the same way as other parameters. We also ran a zero-eccentricity model to the data and find no significant changes in the final parameters.

The residual RVs are consistent with zero after subtracting the best-fit hot Jupiter signal and the long-term trend containing the systemic velocity. We found an RV jitter of $8.1 \pm 3.3$ m/s during the MCMC. We list the results from the fit in Tables \ref{tab: stellar parameters}, \ref{tab: system parameters} and \ref{tab: LD coefficients}.

\subsection{Analysis of Gaia DR2 data}
\label{sec: Analysis of Gaia DR2 data}

Gaia DR2 data can be used to constrain exoplanet system parameters. In our combined fit, we used the Gaia DR2 parallax as an input. As a case study, we discuss the validity of this measurement using Gaia quality indicators and other studies.

There are a number of indications that the Gaia DR2 astrometry for this star is reliable. Its proper motion and parallax from Gaia DR2, Gaia DR1, and Tycho-2 are in full agreement. There is no nearby bright source: the nearest Tycho-2 star is more than 99.9 arcsec away (prox = 999). There is no indication from Gaia DR1 of a long-term curvature of the proper motion which could be caused by a companion (astrometric\_delta\_q = 0.00). The ecliptic latitude of the star ($\beta = 71.28\degree$) is in general good for astrometry, as confirmed by the Gaia DR2 statistics: 219 of the 227 observations (called ``transits'') have been used in the astrometric solution, distributed over 18 visibility periods, far above the minimum number of five periods required.
The mean parallax factor is normal (mean\_varpi\_factor\_al = 0.040), indicating that the astrometric fit should be straightforward, and the astrometric excess noise is zero (astrometric\_excess\_noise = 0.000 mas), which confirms the high quality of the Gaia DR2 astrometry.
One suspicious element is the duplicate source flag (duplicated\_source = 1). However, the number of ``transits'' is similar to that of the nearby faint star (Sec. \ref{sec: Faint nearby star}) which has a duplicate source flag of 0. This precludes that the second, duplicated source identifier is hiding large amounts of data; thus the Gaia DR2 astrometry is based on most data. The astrometric goodness of fit is poor (astrometric\_gof\_al = 9.9804), but the Revised Unit Weight Error (which is a rescaled astrometric quality indicator) is 1.12, which is lower than the threshold of 1.4 that indicates suspicious astrometry. The duplicate source flag likely originates from the partial saturation of this object.

\citet{Lindegren2018, LindegrenPres2018} state that the DR2 error for Gaia parallaxes does not represent the total uncertainty. We increased the uncertainty of the XO-7 parallax using the calibration formula given in slide 17 of \citet{LindegrenPres2018} (see ``Known issues with the Gaia DR2 data''\footnote{\url{https://www.cosmos.esa.int/web/gaia/dr2-known-issues}}). We also considered a systematic offset in the Gaia DR2 parallax of XO-7. \citet{Hall2019} provide a compilation of literature values in addition to their own finding and report a systematic offset around $-50\;\mu$as (see their Fig. 8, Tables 9 and 10). We ran our analysis for two extreme cases: no offset and a $-82\,\pm\,33\;\mu$as offset as found by \citet{Stassun2018}, propagating the uncertainties accordingly. The largest discrepancy between best-fit parameters from the two analyses was a 1$\sigma$ difference in the stellar radius. Because the posterior values do not significantly change after the systematic parallax correction, we used the original Gaia DR2 parallax value and the calibrated uncertainty in our final analysis.

The effective temperature given by Gaia DR2 ($T_{\rm eff} = 5877 \; [5706 \; 6031] \; \rm K$) is lower than the one we measure from high resolution spectroscopy (the values in brackets are the 16th and 84th percentiles of the probability density function). The radius from Gaia DR2 based on the $BP$ and $RP$ magnitudes ($R = 1.58 \; [1.51 \; 1.68] \; \rm R_{\odot}$) is slightly larger than inferred from our fit. Gaia values are useful for ensemble analysis but are not necessarily accurate for single objects. The system radial velocity from Gaia DR2 ($V_{\rm sys} = -12.82 \pm 0.44 \rm \; km\,s^{-1}$) is in excellent agreement with our measurement.

\subsection{Secondary eclipse}
\label{sec: Secondary eclipse}

We do not detect the secondary eclipse in the XO lightcurve. We set an upper limit on its depth $\delta_e$ in the XO bandpass by calculating the noise in the folded lightcurve. We eliminate the in-transit points, fold the lightcurve at $10^5$ different periods ranging from 2 to 3.7 days, split each lightcurve into segments equal to the eclipse duration (assumed to be equal to the transit duration), and calculate the mean flux in each segment. This yields the distribution of flux variations over the timescale of the eclipse. We take the $3\,\sigma$ values of this distribution as the $3\,\sigma$ upper limit on $\delta_e$. We find $\delta_e < 0.00142$ at $3\,\sigma$ in the $R$ band. This limit is too high to constrain the brightness temperature and albedo of the planet's day side.

\subsection{Transit timing variations}
\label{sec: Transit timing variations}

We measure the central times $t_{c}$ of individual transits that were observed during the follow-up campaign. We do not find any correlation pattern between the $t_{c}$ and the period index of the transits, calculated as the number of orbital periods after the first transit. We put an upper limit on the presence of transit timing variations (TTVs) by measuring the standard deviation of the distribution of $t_{c}$: TTVs of XO-7b should be lower than 5 min at 1$\sigma$ (15 min at 3$\sigma$) over the two years of observations. This is consistent with the fact that our RV measurements rule out the presence of companions massive and close enough to induce significant TTVs on shorter timescales. \modif{We note that measuring transit timing variations from ground-based observations that are affected by correlated noise is challenging, as studied by \citet{Carter2009}.}

\section{System parameters}
\label{sec: System parameters}

In this section, we list the parameters of the host star and hot Jupiter as determined by the preceding analysis. We also report the presence of a wide-orbit companion implied by a slope in the radial velocity measurements.

\subsection{Stellar parameters}
\label{sec: Stellar parameters}

The host star is BD+85 317 and is classified as G0V \citep{Pickles2010}, sometimes G2 \citep{Wright2003}. Its parameters are reported in Table~\ref{tab: stellar parameters}. It is bright and circumpolar (within 5\degree of the celestial north pole), which could facilitate follow-up observations from the northern hemisphere. It has a high metallicity that is among the highest for stars harbouring a hot Jupiter. Its age estimated from the best fit MIST isochrone in the $T_{\rm eff}-{\rm log}\,g$ space indicates that it is relatively young. However, several isochrones of disparate ages are in close proximity and provide a good fit as reflected by the large uncertainty (nearly 100\%). Thus, the age should be taken with caution and further study such as activity indicator analysis would be necessary for a better estimate. If indeed the star is young, XO-7 b would be one of the very few hot Jupiters known around young stars.

\begin{table}
\begin{center}
\caption{Parameters of the planet host star.}
\label{tab: stellar parameters}
\begin{tabular}{cccc}
\hline
\hline
   Quantity &  Unit &  Value  & Notes   \\
\hline

Name  &  &  BD+85 317   &  1   \\
RA   & J2000 & 18:29:54.929 &  2   \\
Dec  & J2000 & +85:13:59.58  &   2   \\
$p$  &  mas   &  $4.2419 \pm 0.0215$  &  2  \\   
$d$ & pc & $234.1 \pm 1.2$  &  3  \\
$\mu_{\alpha}$  & $\rm mas\,yr^{-1}$  &  $-15.354 \pm 0.038$  &  2  \\   
$\mu_{\delta}$  & $\rm mas\,yr^{-1}$  &  $24.461 \pm 0.054$  &  2  \\   
$\gamma$  & $\rm km\,s^{-1}$  &  $-12.983 \pm 0.015$  &  7   \\
$B$ & mag  &  $11.23 \pm 0.06$ &   4  \\
$V$ & mag  &  $10.52 \pm 0.04$ &   4  \\
$G$ & mag  &  $10.4575 \pm 0.0004$   &   2   \\
$J$  & mag  & $9.557 \pm 0.024$ &  5   \\
$H$  & mag  & $9.308 \pm 0.030$ &  5   \\
$K$  & mag  & $9.241 \pm 0.024$ &  5   \\
$BP$  & mag  & $10.7795  \pm 0.0008$ &  2   \\
$RP$  & mag  & $10.0087 \pm 0.0010$ &  2   \\
$BP-RP$ & mag  &  $0.7707$  &  2   \\
$AG$ & mag  &  $0.6980 \; [0.5709 \; 0.8174]$  &  2, 8   \\
$E(BP-RP)$ & mag  &  $0.3450 \; [0.2543 \; 0.4170]$  &   2, 8  \\
Sp Type &  & G0V &  6   \\
$T_{\rm eff}$ & K  &  $6250 \pm 100$ &  7  \\
$\rm[Fe/H]$ & dex  &  $0.432 \pm 0.057 $  &  7    \\
log$\,g$ 	& cgs & $4.246 \pm 0.023$   &  7  \\
\vsini 	&  $\rm km\,s^{-1}$  & $6 \pm 1$  &  7  \\
$M$	& M$_{\odot}$	& $1.405 \pm 0.059$	&  7  \\
$R$	& R$_{\odot}$	& $1.480 \pm 0.022$	&  7  \\
Age	& Gyr & $1.18^{+0.98}_{-0.71}$ & 7 \\
${\rm RV\ slope}$ & m ${\rm s}^{-1}$ ${\rm d}^{-1}$ & $-0.148 \pm 0.011$ & 7 \\
\hline  
\end{tabular}
\end{center}
Notes. 1: \citet{Argelander1903}. 2: \citet{Gaia2018}. 3: \citet{BailerJones2018}. 4: \citet{Hog2000}. 5: \citet{Cutri2003}. 6: \citet{Pickles2010}. 7: This work. 8: The values in brackets are the 16th and 84th percentiles of the probability density function.

\end{table}

\subsection{Hot Jupiter parameters}
\label{sec: hot Jupiter parameters}

The hot Jupiter has an orbital period of 2.864 days, a mass of $0.709 \pm 0.034 \; \mj$ and a radius of $1.373 \pm 0.026 \; \rj$ yielding a density of $0.340 \pm 0.027 \; \rm g \, cm^{-3}$. At a distance of $0.04421 \pm 0.00062$ AU from its star, the planet has an equilibrium temperature of $1743 \pm 23 \; \rm K$ assuming a zero albedo. The orbit is consistent with being circular. The parameters of the hot Jupiter are reported in Table \ref{tab: system parameters} and the best-fit limb darkening coefficients for each band are reported in Table \ref{tab: LD coefficients}.

\begin{deluxetable*}{lccccccccccccccccccccccccccccccccc}
\tablecaption{Median values and 68\% confidence intervals for XO-7 b.\label{tab: system parameters}}
\tablehead{\colhead{~~~Quantity} & \colhead{Unit} & \multicolumn{32}{c}{~~~~~~~~~~~~Value~~~~~~~~~~~~~}}
\startdata
~~~~$P$\dotfill &Period (days)\dotfill &$2.8641424\pm0.0000043$\\
~~~~$R_P$\dotfill &Radius (\rj)\dotfill &$1.373\pm0.026$\\
~~~~$T_C$\dotfill &Time of conjunction (\bjdtdb)\dotfill &$2457917.47503\pm0.00045$\\
~~~~$a$\dotfill &Semi-major axis (AU)\dotfill &$0.04421\pm0.00062$\\
~~~~$i$\dotfill &Inclination (Degrees)\dotfill &$83.45\pm0.29$\\
~~~~$e$\dotfill &Eccentricity \dotfill &$0.038\pm0.033$\\
~~~~$T_{eq}$\dotfill &Equilibrium temperature (K)\dotfill &$1743\pm23$\\
~~~~$M_P$\dotfill &Mass (\mj)\dotfill &$0.709\pm0.034$\\
~~~~$K$\dotfill &RV semi-amplitude (m/s)\dotfill &$80.5\pm3.2$\\
~~~~$R_P/R_*$\dotfill &Radius of planet in stellar radii \dotfill &$0.09532\pm0.00093$\\
~~~~$a/R_*$\dotfill &Semi-major axis in stellar radii \dotfill &$6.43\pm0.14$\\
~~~~$\delta$\dotfill &Transit depth (fraction)\dotfill &$0.00909\pm0.00018$\\
~~~~$\tau$\dotfill &Ingress/egress transit duration (days)\dotfill &$0.0190\pm0.0015$\\
~~~~$T_{14}$\dotfill &Total transit duration (days)\dotfill &$0.1155\pm0.0014$\\
~~~~$T_{FWHM}$\dotfill &FWHM transit duration (days)\dotfill &$0.09655\pm0.00074$\\
~~~~$b$\dotfill &Transit impact parameter \dotfill &$0.709\pm0.023$\\
~~~~$\delta_{S,3.6}$\dotfill &Blackbody eclipse depth at 3.6 $\mu$m (ppm)\dotfill &$898\pm29$\\
~~~~$\delta_{S,4.5}$\dotfill &Blackbody eclipse depth at 4.5 $\mu$m (ppm)\dotfill &$1150\pm34$\\
~~~~$\rho_P$\dotfill &Density (cgs)\dotfill &$0.340\pm0.027$\\
~~~~$logg_P$\dotfill &Surface gravity \dotfill &$2.970\pm0.028$\\
~~~~$\Theta$\dotfill &Safronov Number \dotfill &$0.0325\pm0.0014$\\
~~~~$\fave$\dotfill &Incident Flux (\fluxcgs)\dotfill &$2.09\pm0.11$\\
~~~~$T_S$\dotfill &Time of eclipse (\bjdtdb)\dotfill &$2457918.900\pm0.029$\\
~~~~$M_P/M_*$\dotfill &Mass ratio \dotfill &$0.000482\pm0.000021$\\
~~~~$P_T$\dotfill &A priori non-grazing transit prob \dotfill &$0.1457\pm0.0073$\\
~~~~$P_{T,G}$\dotfill &A priori transit prob \dotfill &$0.1764\pm0.0087$\\
\enddata
\end{deluxetable*}

\begin{deluxetable}{lccccccccccccccccccccccccccccccccc}
\tablecaption{Best-fit limb darkening coefficients using a quadratic law.\label{tab: LD coefficients}}
\tablehead{\colhead{Band} & \colhead{$u_1$} & \colhead{$u_2$}}
\startdata
B &$0.586\pm0.031$ & $0.213\pm0.026$ \\
V &$0.404\pm0.024$ & $0.296\pm0.019$ \\
g' &$0.504\pm0.054$ & $0.250\pm0.052$ \\
R &$0.317\pm0.024$ & $0.321\pm0.020$ \\
r' &$0.340\pm0.052$ & $0.313\pm0.050$ \\
i' &$0.252\pm0.032$ & $0.312\pm0.029$ \\
C &$0.352\pm0.029$ & $0.311\pm0.026$ \\
\enddata
\end{deluxetable}

\subsection{Wide orbit companion}
\label{sec: Wide orbit companion}

The radial velocity measurements show a linear trend in addition to the radial velocities induced by the hot Jupiter, indicating the presence of a wide orbit companion. The secular change is 100 m/s over two years and no curvature is apparent in the data. Thus, the minimum orbital period is four years in the case of a circular orbit and two years for a very eccentric orbit. Assuming a circular orbit, we derive a minimum mass of $4 \, \mj$ for the companion, which could be a planet, a brown dwarf or a star. This system is still under monitoring to characterize the long period companion.

\section{Discussion}
\label{sec: Discussion}

\subsection{Prospects for atmospheric characterization}

\begin{figure}[htbp]
   \centering
   \includegraphics[width=8cm]{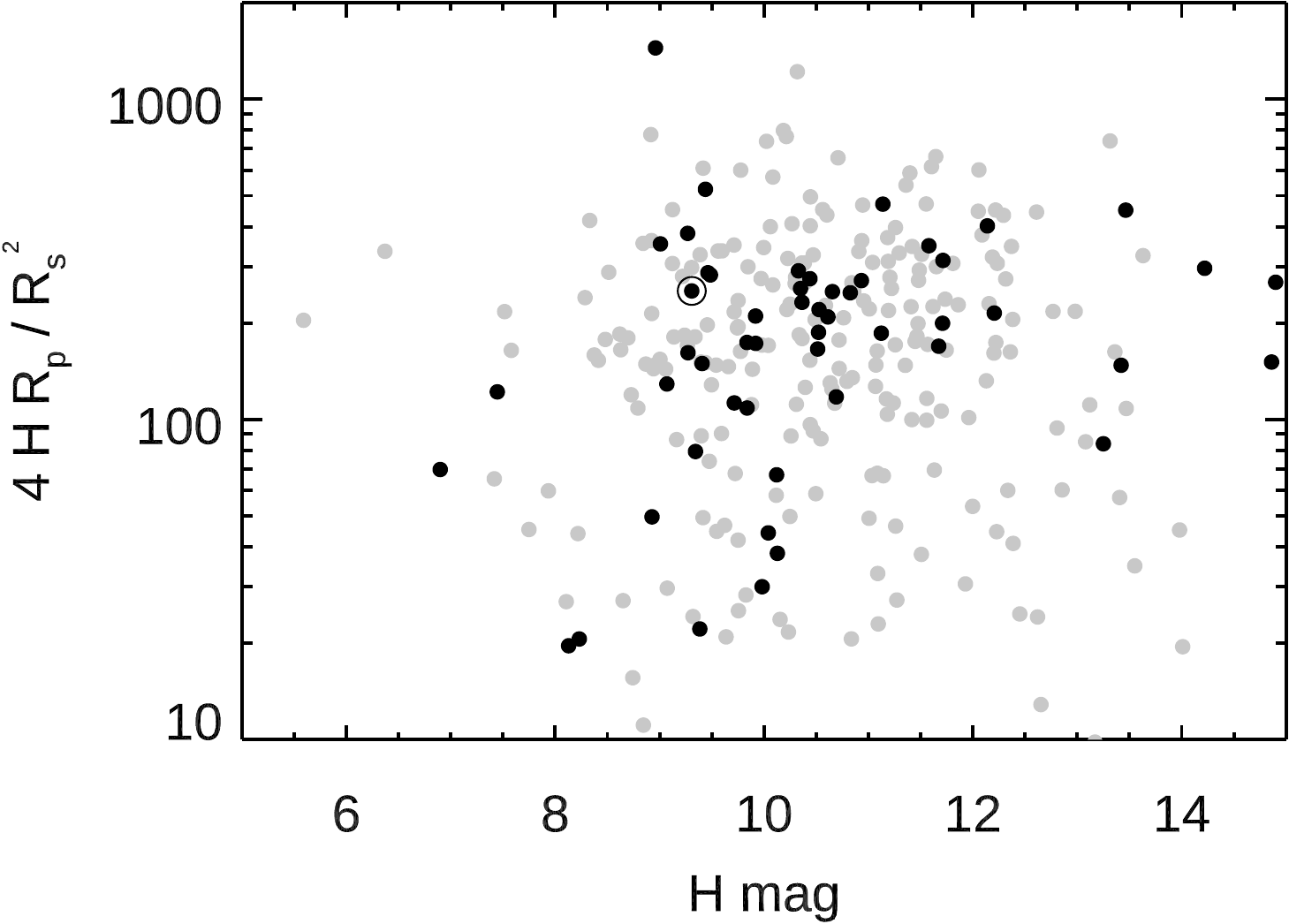}
     \caption{\modif{Estimated planetary atmospheric transmission signal as a function of host star $H$ magnitude for known transiting hot Jupiters (defined as $0.3 \; \mj < M_{\rm p} < 13 \; \mj$ and $P < 20$~d). Host stars visible by \jwst more than 200 days per year are shown in black. XO-7 is highlighted by a black open circle. Data are from exoplanet.eu, simbad.u-strasbg.fr, and the \jwst General Target Visibility Tool.}}
   \label{fig: vmag scaleheight}
\end{figure}

XO-7 b is an inflated hot Jupiter and is moderately hot. Its large atmospheric scale height ($H$ = 671~km) combined with the brightness of the host star makes it well suited to atmospheric characterization. It is among the 25 known transiting hot Jupiters with an atmospheric scale height larger than 500~km and a host star brighter than magnitude 11 in the $V$ band. Assuming an absorption spanning two scale heights and estimating the amplitude of the transmission signal by $2\,H\times2\,R_p/R_\star^2$, we expect a signal of 250 ppm, which could be detected with \hst (Fig. \ref{fig: vmag scaleheight}). Thus, XO-7 b is a valuable target to investigate the atmospheric properties of moderately irradiated close-in gas giant planets.

Among known transiting hot Jupiters with a mass, radius, and equilibrium temperature within 20\% of those of XO-7 b, two of them have been observed in spectroscopy with \hst STIS and/or WFC3: \mbox{HD 209458 b} and \mbox{HAT-P-13 b}. Atomic and molecular species have been detected in the atmosphere of \mbox{HD 209458 b} including water vapour signatures around 1.4 $\mu$m with an amplitude of 200 ppm, which is about twice smaller than expected for a clear, solar composition atmosphere, and indicates extra absorption by haze and/or dust \citep{Deming2013, Sing2016}. This low amplitude might also be due to a depletion in oxygen compared to solar abundance \citep{Madhu2014} although this explanation is not favored at the moment. The \mbox{HAT-P-13 b} WFC3 observations have not been published but are available on the MAST archive. Observing XO-7~b in transit spectroscopy would test if the atmospheric properties measured for HD 209458 b are also valid for a planet with similar characteristics and would help constrain hot Jupiter atmosphere models.

The host star has a high ecliptic latitude ($\beta = 71.28\degree$) and will be visible during a continuous period of 212 days with \jwst. In addition, it is bright but will not saturate the \jwst detectors in time series spectroscopic observation modes (except in the NIRSpec PRISM/CLEAR configuration) and there is no surrounding star that could contaminate the spectrum (the faint nearby star is far enough away). Thus, it is an excellent target for \jwst. It will also be a good target for ARIEL \modif{for similar reasons: it will be visible continuously, it is bright, and the presence of the companion makes it valuable as part of a larger sample to investigate connections between formation and migration mechanisms and atmospheric compositions. Thus, it would be a good addition to the ARIEL target list.}

\subsection{Prospects from \tess follow-up}

The transiting extrasolar planet XO-7 b reported here is probably
the last of the series discovered by the XO project \citep{McCullough2005}. This section addresses some of the ways in which data
from a ground-based survey such as XO differ from data from the
\tess mission.\footnote{In this section, we assume that \tess will
complete its two-year survey nominally (with shifted sectors 14, 15 and 16 in Cycle 2), and except
where noted, extensions of the \tess primary mission will not be
required to obtain the results discussed.} Comparison of XO data
for XO-7 b and \tess data for a similar planet (WASP-126b) demonstrates
that for discovering transiting planets, \tess data will be far
superior. \tess will observe XO-7 as TIC 268403451 in camera 3 during
its Sectors 18-20, between November 2019 and January 2020. The \tess
light curve will consist of nearly-continuous monitoring of XO-7
for three months at 2-minute cadence.  For comparison, we selected
an exoplanet candidate (TIC 25155310; TOI-114.01) with similar
properties to XO-7/XO-7 b from the many candidates already reported
by \tess. Like XO-7 b, the example that we selected happens to be a
re-discovery by \tess of a planet, WASP-126b, discovered by a
ground-based survey. Whereas XO-7 b transits a V=10.5 mag G0 V star
every 2.9 days, WASP-126b transits a V=10.8 mag G2 V star every 3.3
days \citep{Maxted2016}.  We obtained a \tess data validation
report for WASP-126b from MAST\footnote{\scriptsize{\url{https://archive.stsci.edu/prepds/tess-data-alerts/\#dataaccess}}}
that includes 21 transits observed during \tess sectors 1-3.
WASP-126b's transit depth is reported as $7182\pm39$ ppm.

Detection of the secondary eclipse of XO-7 b would permit a measure
of the superposition of reflected starlight and emission from the
planet's illuminated side. It would also confirm whether the orbital
eccentricity is indeed close to zero.  Unfortunately,
the prospects are very poor for detecting the secondary eclipse of
XO-7 b with \tess data.  Based upon the reported uncertainty of the
depth of WASP-126b's transit, a secondary eclipse of similar duration
would have been marginally detected by \tess at $3\sigma$ if its
depth were 117 ppm.  However, if the geometric albedo of XO-7 b is
typical of hot Jupiters, e.g. 0.1, then its secondary eclipse depth
due solely to reflected light is expected to be 21 ppm \citep[][Eq. 4]{Sheets2017}.
While thermal emission from a hot Jupiter can contribute to the
depth of its secondary eclipse, even within the 600-1000 nm bandpasses
of \tess or Kepler, \citet[][Table 4]{Sheets2017} report a median
of 35 ppm and a maximum of 91 ppm for the depths of secondary
eclipses of 14 hot Jupiters.  Because those two estimates of secondary
eclipse depths (21 ppm and 35 ppm) are much smaller than the 117
ppm estimate of a marginal ($3\sigma$) detection of such an eclipse
with \tess data similar to that expected for XO-7, we expect that \tess
will not detect XO-7 b's secondary eclipse.

Similarly, a search of \tess data of XO-7 could turn up transits of
a planet smaller in radius than XO-7 b. However, such a search is
likely to be fruitless because such companions to hot Jupiters are
rare \citep{Steffen2012}.  Likewise, measuring XO-7's mean density
or age from asteroseismology will not be possible with \tess data:
XO-7 is a few magnitudes too faint \citep[cf.][Fig. 12a]{Campante2016}.  \tess photometry may enable measuring the rotation
period of the star XO-7, and comparing XO-7's rotation period and
its radius with its spectroscopically determined $v \, {\rm sin}(i)$ may yield
the star's spin axis inclination ($i$) for comparison with the
projected angle of orbital obliquity of XO-7 b inferred from the
Rossiter-McLaughlin effect \citep{Holt1893, Rossiter1924, McLaughlin1924}. For stars hotter than approximately
6250 K, the pole of the planet's orbit could be preferentially misaligned with
respect to the stellar spin axis \citep{Winn2010a}; XO-7's effective
temperature ($6250 \pm 100$ K; Table \ref{tab: stellar parameters}) places it at the threshold.

\subsection{Follow-up of the wide orbit companion}

The secular trend in XO-7's radial velocities indicates an unseen
companion, either a planet, brown dwarf, or star.
\citet{Ngo2016} estimate that the hosts of hot Jupiters
have stellar companions with separations less than 50 AU in $3.9^{+4.6}_{-2.0}$\%
of their sample.  \citet{Knutson2014} estimate that $27\pm6$\% of hot Jupiters
have a planetary companion in the range of mass $1-13 \: \rm M_J$ and semi-major axis
$1-10 \: \rm AU$. In absolute value, the slope of XO-7's radial velocities, $-0.148 \pm 0.001 \: \rm m \,s^{-1} \,{day}^{-1}$, is 1.5 times
larger than the maximum slope ($-0.097 \pm 0.023 \: \rm m \,s^{-1} \,{day}^{-1}$)
exhibited in a sample of 51 hot Jupiters observed by \citet{Knutson2014}, typically for at least five years each.\footnote{This statement ignores
the three systems for which inflexions enabled a two-planet solution: HATP-17, WASP-8, WASP-34.}
The slope of XO-7's radial velocities implies the
following relationship between the unseen companion's mass ($M_c$,
in solar units) and its angular separation from XO-7 ($\theta$, in
arc seconds) \citep[][Eq. 1]{Knutson2014}:
\begin{equation}
M_c = 40 \: \theta^2,
\end{equation}
which implies that the companion must be hidden within a nominal
ground-based seeing disk (2\arcsec\ FWHM), otherwise it would be
much more massive than XO-7 and would dominate the light.  If it
is a low-mass star (0.1 to 0.4 $M_{\odot}$), then its separation
is 0.05\arcsec to 0.1\arcsec, or 12 AU to 24 AU, and could be
revealed, with adaptive optics as a main-sequence M star, respectively
7 to 4 magnitudes fainter than XO-7 in K band \citep[][Fig.~1]{Delfosse2000}. 

Regardless of whether it is a planet, brown dwarf, or
a star, if its orbit is not very elliptical, then its orbital
period must be measured in years, otherwise the trend of radial
velocities would show some degree of curvature. For example, a $M {\rm sin}(i) = 5 \, \rm M_J$
companion in a five-year circular orbit is consistent with the
radial velocities measured to date. In that case, Gaia astrometry
would be able to detect at $\sim 5\, \sigma/{\rm sin}(i)$ the $\sim 50/{\rm sin}(i)$
micro-arcsecond astrometric wobble of XO-7 induced by such a
companion \citep{Perryman2014}.
Astrometric orbits from Gaia will be available in DR4. Combining the astrometric orbit of XO-7 with our radial velocity measurements will allow us to reconstruct the orbits in three dimensions and measure the mass of the wide orbit companion.
If it is stellar, the companion may also be detected in photometry by Gaia  \citep{deBruijne2015}. Companions at such small separations are very incomplete in Gaia DR2 and need special ground processing \citep{Arenou2018}; they will be available in Gaia DR4.

\subsection{Migration and orbit of the hot Jupiter}

Hot Jupiters are thought to form beyond the water ice line and migrate inwards to reach close-in orbits. Two mechanisms have been proposed: disk or high eccentricity migration. Disk migration should yield orbits that are circular and contained in the plane perpendicular to the star rotation axis, and can go on until the planet reaches the Roche limit $a_{Roche}$. In contrast, high eccentricity migration should yield a wide range of obliquities and the planet is expected to reach a circular orbit at a distance almost exactly $2\,a_{Roche}$ \citep{Faber2005}. The orbit may also retain some eccentricity depending on the circularization timescale. In this process, the initial eccentricity can originate from scattering interactions with another massive companion.
Although no correlations have been found between hot Jupiter orbital parameters and the presence of distant companions \citep{Knutson2014}, we place XO-7 in the context of these mechanisms.
XO-7 has an outer companion of at least $4 \; \rm M_J$ and the hot Jupiter appears to have a nearly circular orbit. We calculate the Roche limit of the star -- hot Jupiter system as $a_{Roche} = 2.7 \, R_p \, (M_\star/M_p)^{1/3}$ and find $a_{Roche} = 0.023 \; \rm AU$. Interestingly, the semi-major axis is almost twice the Roche limit ($a/a_{Roche} = 1.95$). This is consistent with expectations for high eccentricity migration (although disk migration is not ruled out). As illustrated in Fig. 8 of \citet{Sarkis2018}, almost all transiting hot Jupiters with a massive outer companion have $a/a_{Roche} > 2$. One exception has $a/a_{Roche}$ just below 2, as for XO-7 b. Overall, this supports high eccentricity migration for hot Jupiters in these systems.

If the planet-planet scattering mechanism played a role in the formation of this system, then we can expect a non-zero obliquity for the hot Jupiter. The wide orbit companion may also affect the hot Jupiter's orbit via the Lidov-Kozai mechanism \citep{Lidov1962, Kozai1962}, which has been proposed to explain the observed sky-projected obliquities of hot Jupiters \citep{Winn2010a, Schlaufman2010, Albrecht2012, Dawson2014, Winn2015}. Our attempts to observe the Rossiter-McLaughlin effect during a transit of XO-7 b to measure its obliquity have been unsuccessful so far, but we plan to make this observation in the near future. This will bring another clue to understand the formation, migration, and architecture of the XO-7 system.

\subsection{Similarities between XO-7 and HAT-P-13}

The XO-7 system has striking similarities with HAT-P-13 \citep{Bakos2009, Winn2010c}. The two hot Jupiters XO-7 b and HAT-P-13 b have, respectively, similar periods (2.86 and 2.92 d), masses (0.71 and 0.85 \mj), radii (1.37 and 1.27 \rj), semi-major axes (0.044 and 0.043 AU), and equilibrium temperatures (1747 and 1653 K); the host stars XO-7 and HAT-P-13 are metal-rich G dwarfs with similar metallicities (0.43 and 0.41 dex), and both have a second companion on a wide orbit. HAT-P-13 c is well characterized ($P = 446.27 \pm 0.22 \; \rm d$, $ M\,{\rm sin}\,i = 14.28 \pm 0.28 \; \rm M_J$, $e = 0.6616 \pm 0.0054$) whereas XO-7 is still under monitoring to characterize that companion. HAT-P-13 has a third outer companion revealed by a linear trend in radial velocities \citep{Winn2010c, Knutson2014}. XO-7 b and HAT-P-13 b have radius ratios $a/a_{Roche}$ of 1.96 and 2.27 and both have a nearly zero eccentricity ($0.036\pm0.032$ and $0.0133 \pm 0.0041$). These systems have different ages ($1.12^{+0.94}_{-0.66}$ and $5.0^{+2.5}_{-0.7}$ Gyr). It would be interesting to investigate if they could have formed in the same way. Also, comparing the XUV emission of the stars for example with the He I 1.08 $\mu$m absorption feature of the hot Jupiter's atmospheres with high-resolution spectroscopy could provide clues on evaporation scenarios of these atmospheres \citep[\textit{e.g.}][]{Allart2018, Nortmann2018}. Finally, both stars are metal rich and host at least two massive companions, and it is known that giant planet formation is correlated with stellar metallicity \citep{Fischer2005}. On the other hand, no statistically significant correlation between the frequency of long-period companions and stellar metallicity has been found in hot Jupiter systems \citep{Knutson2014}. Thus, how to interpret these high metallicities in the context of the formation of these systems remains an open question.

\section{Conclusion}
\label{sec: Conclusion}

We report the discovery of the transiting hot Jupiter XO-7 b orbiting a main sequence G0 star. Its bright host star and large atmospheric scale height make it well suited to atmospheric characterization. Its physical properties are close to those of HD 209458 b, which has been extensively characterized, and even closer to HAT-P-13 b. Inferring whether their atmospheres also have similar properties would help constrain hot Jupiter atmosphere models. The object is circumpolar which could facilitate follow-up observations from the ground. We detect the presence of a more massive, wide orbit companion with a period of at least a few years. Radial velocity monitoring is underway to determine whether this companion is a planet, a brown dwarf, or a low mass star. In addition, the astrometric motion of the host star caused by that companion should be detectable by Gaia and available in DR4. Combining these measurements will yield the orbits in three dimensions. If it is a low mass star, that companion may also be seen in adaptive optics imaging. \tess photometry of XO-7 will yield improved parameters of the hot Jupiter and host star and may provide the star's rotation period, from which we could determine its spin axis inclination. Measuring the Rossiter-McLaughlin effect in radial velocities will yield the hot Jupiter's obliquity and a potential link with the wide orbit companion may be investigated. Finally, the hot Jupiter orbital parameters and the presence of a wide orbit companion are consistent with expectations for a high eccentricity migration mechanism. Thus, this discovery is valuable to investigate the formation and evolution of hot Jupiter systems.

\acknowledgments
We thank the anonymous referee for helpful thoughts and critique that strengthened our paper. We also thank the editor for overseeing our submission.
The XO project is supported by NASA grant NNX10AG30G.
I.R., F.V. and E.H. acknowledge support by the Spanish Ministry for Science, Innovation and Universities (MCIU) and the Fondo Europeo de Desarrollo Regional (FEDER) through grant ESP2016- 80435-C2-1-R, as well as the support of the Generalitat de Catalunya/CERCA programme. The Joan Or\'{o} Telescope (TJO) of the Montsec Astronomical Observatory (OAdM) is owned by the Generalitat de Catalunya and operated by the Institute for Space Studies of Catalonia (IEEC).
NCS was supported by FCT - Funda\c{c}\~{a}o para a Ci\^encia e a Tecnologia through national funds and by FEDER - Fundo Europeu de Desenvolvimento Regional through COMPETE2020 - Programa Operacional Competitividade e Internacionaliza\c{c}\~{a}o by these grants: UID/FIS/04434/2019; PTDC/FIS-AST/28953/2017 \& POCI-01-0145-FEDER-028953 and PTDC/FIS-AST/32113/2017 \& POCI-01-0145-FEDER-032113.
HPO acknowledges support from Centre National d'Etudes Spatiales (CNES) grant 131425-PLATO. This research made use of Photutils, an Astropy package for
detection and photometry of astronomical sources \citep{bradley2019}. This research has made use of the Exoplanet Orbit Database and the Exoplanet Data Explorer at exoplanets.org, the Extrasolar Planets Encyclopaedia at exoplanet.eu, and the SIMBAD and VizieR databases at simbad.u-strasbg.fr/simbad/ and http://vizier.u-strasbg.fr/viz-bin/VizieR.

%

\vspace{5mm}
\facilities{OHP(SOPHIE), LCOGT:0.4m, Montsec Astronomical Observatory:Joan Or\'o Telescope, Nice Observatory:Schaumasse.}


\software{astrometry.net \citep{Lang2010}, 
          Stellar Photometry Software \citep{Janes1993}, 
          \modif{IRIS \citep{Buil2005},
          ICAT \citep{Colome2006},
          AstroImageJ \citep{Collins2017},
          Photutils \citep{bradley2019},
          JKTLD \citep{Southworth2015}, 
          EXOFASTv2 \citep{Eastman2019}}.   
          }

\bibliography{biblio}

\appendix

\section{SOPHIE radial velocities}

\label{tab: xo7 RVs}

Radial velocities of the host star XO-7 measured along the orbit using the SOPHIE spectrograph at the Observatoire de Haute-Provence, France, between July 23, 2016 and July 4, 2018. \\

\begin{minipage}{0.5\textwidth}
\begin{center}

\begin{tabular}{cccc}

\hline\hline
Reduced  & Orbital   & RV  & $\sigma$   \\
BJD & phase &  [$\rm km\,s^{-1}$] &  [$\rm km\,s^{-1}$]  \\
\hline
57593.3798	&      0.84405	&	-12.885	&	0.018  \\
57623.3312	&      0.30141	&	-13.002	&	0.013  \\
57624.3611	&      0.66100	&	-12.852	&	0.013  \\
57627.3535	&      0.70577	&	-12.853	&	0.014  \\
57628.3640	&      0.05859	&	-12.954	&	0.013  \\
57659.4151	&      0.89990	&	-12.898	&	0.014  \\
57660.3593	&      0.22957	&	-13.025	&	0.013  \\
57661.4453	&      0.60874	&	-12.873	&	0.013  \\
57681.4732	&      0.60136	&	-12.913	&	0.020  \\
57682.3493	&      0.90725	&	-12.897	&	0.010  \\
57719.4319	&      0.85443	&	-12.846	&	0.013  \\
57744.2203	&      0.50916	&	-12.946	&	0.013  \\
57745.2956	&      0.88460	&	-12.906	&	0.013  \\
57746.3182	&      0.24163	&	-13.020	&	0.013  \\
57879.5980	&      0.77553	&	-12.925	&	0.017  \\
57907.5616	&      0.53886	&	-12.971	&	0.013  \\
57908.5474	&      0.88305	&	-12.935	&	0.013  \\
57909.5485	&      0.23258	&	-13.043	&	0.014  \\
57910.4944	&      0.56283	&	-12.933	&	0.012  \\
57911.5923	&      0.94616	&	-12.967	&	0.012  \\
57938.5932	&      0.37337	&	-13.037	&	0.011  \\
57941.5753	&      0.41456	&	-13.037	&	0.008  \\
57956.3772	&      0.58256	&	-12.963	&	0.026  \\
57974.4250	&      0.88384	&	-12.927	&	0.025  \\
57976.3684	&      0.56237	&	-12.949	&	0.013  \\
\hline\hline

\end{tabular}
\end{center}
\end{minipage}
\quad
\begin{minipage}{0.5\textwidth}
\begin{center}
\begin{tabular}{cccc}

\hline\hline
Reduced  & Orbital   & RV  & $\sigma$   \\
BJD & phase &  [$\rm km\,s^{-1}$] &  [$\rm km\,s^{-1}$]  \\
\hline
57978.3666	&      0.26003	&	-13.075	&	0.013  \\
57979.3836	&      0.61511	&	-12.956	&	0.013  \\
57987.3315	&      0.39008	&	-13.035	&	0.013  \\
57988.3711	&      0.75305	&	-12.922	&	0.011  \\
57989.3676	&      0.10097	&	-13.056	&	0.014  \\
58004.6152	&      0.42458	&	-13.051	&	0.020  \\
58007.4696	&      0.42118	&	-13.029	&	0.020  \\
58008.4712	&      0.77088	&	-12.920	&	0.023  \\
58036.2733	&      0.47783	&	-13.008	&	0.013  \\
58038.3084	&      0.18838	&	-13.098	&	0.013  \\
58039.2630	&      0.52167	&	-12.977	&	0.013  \\
58054.2434	&      0.75199	&	-12.911	&	0.017  \\
58057.2494	&      0.80152	&	-12.925	&	0.013  \\
58201.6742	&      0.22663	&	-13.093	&	0.012  \\
58203.6295	&      0.90931	&	-12.970	&	0.014  \\
58230.6347	&      0.33803	&	-13.108	&	0.017  \\
58231.6159	&      0.68061	&	-12.938	&	0.013  \\
58233.6379	&      0.38658	&	-13.075	&	0.013  \\
58247.6194	&      0.26814	&	-13.089	&	0.013  \\
58286.4787	&      0.83565	&	-12.951	&	0.014  \\
58299.5470	&      0.39837	&	-13.056	&	0.018  \\
58300.5370	&      0.74402	&	-12.966	&	0.013  \\
58302.4766	&      0.42122	&	-13.064	&	0.013  \\
58303.5649	&      0.80120	&	-12.944	&	0.013  \\
\hline\hline

\end{tabular}

\vspace{2mm}
Note. The orbital phase is 0 at mid-transit.
\end{center}

\end{minipage}

\vspace{-2mm}



\end{document}